# VCSEL-based PAM-4 transmission system emulator: A data-driven deep learning perspective

**Stavros Deligiannidis,[1*] Nikos Argyris,[2] Stefanos Dris,[2] Dimitris Kalavrouziotis,[2] Paraskevas Bakopoulos,[2] Charis Mesaritakis,[3] and Adonis Bogris[1]**

[1]*Department of Informatics and Computer Engineering, University of West Attica, Egaleo 12243, Greece*
[2]*NVIDIA, Ermou 56, Athens 10563, Greece*
[3]*Department of Biomedical Engineering, University of West Attica, Egaleo 12243, Greece*
**sdeligiannid@uniwa.gr*

**Abstract:** We demonstrate a data-driven framework for emulating high-speed VCSEL-based 4-level Pulse Amplitude Modulation (PAM-4) optical interconnects using bidirectional Long Short-Term Memory (Bi-LSTM) networks. Unlike conventional rate-equation models, which are computationally intensive and often require difficult parameter tuning, our approach utilizes experimental waveforms to learn the end-to-end system dynamics. By employing transfer learning and weight interpolation, we extend the model to new operating regimes with a 20-fold reduction in computation time compared to independent training, while maintaining normalized mean squared error below 0.04. This emulator provides a rapid, accurate tool for the design and optimization of short-reach optical links.

## 1. Introduction

The design and performance evaluation of photonic and opto-electronic systems are crucial for advancing technologies in data communications, sensing, and imaging. These systems rely on high-speed, energy-efficient components that must operate reliably under varying conditions. Traditionally, numerical simulations and iterative design adjustments have been the primary tools for optimizing photonic devices. However, these methods can be computationally demanding and struggle to fully capture the complex interactions inherent in real-world implementations [1].

Vertical-cavity surface-emitting lasers (VCSELs) serve as key components in data communication networks (DCNs), providing energy-efficient, high-speed optical links for short-reach interconnects. Their widespread adoption is driven by their compact design, low fabrication costs, and ability to be directly modulated at high data rates. However, VCSEL-based transmission systems exhibit complex nonlinear behavior influenced by bandwidth limitations, carrier dynamics, and modal dispersion. Accurately predicting the interplay of these effects is crucial for optimizing their performance across different data rates and operating conditions [2].

Traditionally, numerical simulations and rate-equation-based models have been used to characterize VCSEL dynamics. While effective within limited operating ranges, these methods rely on sequential time-step integration. Consequently, they become computationally heavy for large-scale system optimization (e.g., sweeping bias currents, temperatures, and fiber lengths) [3]. Furthermore, they often require extensive parameter tuning to match experimental behavior, an approach that is both time-consuming and resource-intensive [4].

To address this engineering bottleneck, neural network-based nonlinear models have emerged for emulating photonic systems. These data-driven approaches offer a more efficient and potentially more accurate alternative to traditional physical models [5]. Recent literature has shown significant progress in this area. For instance, Bidirectional Long Short-Term

Memory (Bi-LSTM) architectures have been used for directly modulated lasers (DML) system identification based on numerical simulations [6], while experimental end-to-end LSTM frameworks have been proposed for optimizing specific Intensity Modulation/Direct Detection (IM/DD) transmission links [7]. However, intra-data center communication is dominated by VCSEL-based active optical cables using multimode fibers. The fast emulation of the performance of an active optical cable is significant in order to know in advance the limits and potential of each transceiver in a densely populated data center. Furthermore, 850 nm VCSELs exhibit fundamentally different dynamics than standard single-mode DMLs, heavily influenced by multi-mode competition, spatial hole burning [8], and distinct RC parasitic constraints [9].

Building upon our initial proof-of-concept demonstration presented at the CLEO 2023 conference [10], which established one of the earliest experimental validations of data-driven VCSEL emulation, this study comprehensively expands the framework. We employ a Bi-LSTM [11,12] to emulate the dynamics of a VCSEL-based PAM-4 transmitter, which has been experimentally characterized. Compared to the traditional rate equation model, which only emulates the VCSEL dynamics in a fully serial process, our approach provides an end-to-end emulation of the entire transmission system taking into account all imperfections and impairments of the transmitter, multi-mode fiber and receiver. Moreover, our method is significantly faster as it can be fully parallelized on a Graphics Processing Unit (GPU), unlocking substantial computational efficiency.

Crucially, to overcome the limitations of training discrete models for individual setups, we introduce a robust transfer learning and weight interpolation strategy. We utilize transfer learning to extend learned weights to new operating regimes, achieving up to a 16-fold reduction in processing time. Furthermore, by adopting an adaptation strategy inspired by reservoir computing [13], where the inner recurrent core is frozen and only the outermost boundary layers are fine-tuned , we achieve a 20-fold reduction in training time. Most notably, we demonstrate that the network weights evolve linearly between fine-tuned states. This enables the derivation of highly accurate models for intermediate continuous operating regimes strictly through mathematical weight interpolation, completely bypassing the need for additional data or retraining. The application of these techniques has yielded very low normalized mean squared error (NMSE) values (0.02–0.04) across all operating conditions, demonstrating their immense potential as a rapid prototyping tool for state-of-the-art transceivers in data center networks.

## 2. Experimental Setup

The experimental setup used to model the dynamic behavior of a VCSEL-based PAM-4 transmitter comprises a 106.25 Gb/s optical transmission system, as depicted in Fig. 1. The system utilizes a high-speed 850 nm multimode VCSEL module (VIS VM100-850M, V-I Systems). This specific transmitter features a modulation bandwidth that supports data rates up to approximately 112 Gb/s for PAM-4 transmissions, providing the necessary capacity for our 53.125 Gbaud (106.25 Gb/s) experimental characterization. The VCSEL is directly modulated with a pseudorandom binary sequence generated by a Mersenne Twister with a period of $2^{19937}$-1. The generated signal is transmitted over a multimode fiber (MMF) and detected by a 30 GHz photodetector (PD), enabling high-speed optical-to-electrical conversion. The bit pattern generator (BPG), in combination with a digital-to-analog converter (DAC), provides a 6-bit resolution.

A 4-tap feed-forward equalizer (FFE) integrated in the DAC was used to compensate for linear impairments arising from the limited VCSEL modulation bandwidth and its electrical interface with the DAC output stage. The tap coefficients were optimized in a back-to-back configuration (~5 m MMF) using an optical sampling module connected to a digital sampling oscilloscope. A Least Mean Squares (LMS) routine was applied to minimize the mean-square error of the recovered PAM-4 levels, and the resulting coefficients were then fixed for all subsequent measurements. This static 4-tap FFE provides the maximum achievable Signal-to-Noise Ratio (SNR) at the transmitter output prior to launching the signal into the MMF link.

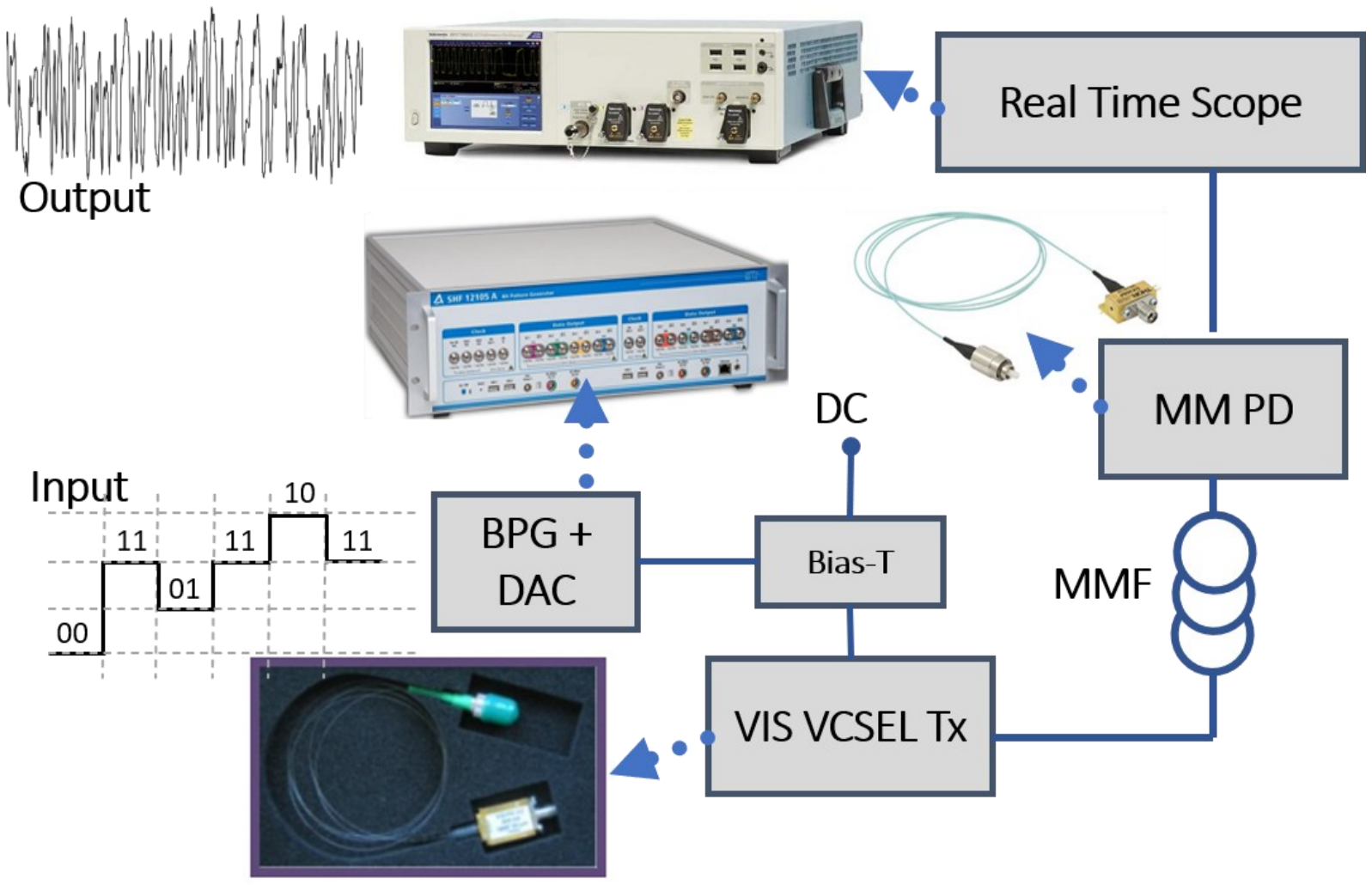

Fig. 1 Experimental setup of the VCSEL PAM-4 optical system

While the FFE can support symbol rates up to 60 Gbaud, it is configured here for 53.125 Gbaud to match the experimental conditions, enhancing signal integrity and minimizing inter-symbol interference. The DC voltage of the Bias-T is used to control the operating regimes of the VCSEL. The electrical SNR of the driving signal in our setup was 25-30 dB. The dominant noise sources were the VCSEL relative intensity noise (-138 dB/Hz) and the photodiode equivalent noise (32 pW/√Hz), ensuring operation in a realistically low-noise regime typical of modern short-reach VCSEL links. Signal monitoring and capturing are performed using a real-time oscilloscope with a 33 GHz analog bandwidth and a sampling rate of 100 GSa/s, ensuring accurate acquisition of the high-speed waveforms.

It is important to emphasize that the multimode (MM) operation of the device significantly increases the complexity and the magnitude of the nonlinear response compared to generic single-mode edge-emitting Directly Modulated Lasers (e.g., 1550 nm Distributed Feedback (DFB) lasers). While the fundamental photon-carrier interaction is conceptually similar, VCSELs support multiple transverse modes. The complex mode competition and spatial hole burning among these modes introduce unique dynamic nonlinearities that standard single-mode rate equations struggle to capture [8]. Furthermore, VCSEL modulation dynamics are heavily influenced by the high RC parasitics inherent to their Distributed Bragg Reflector (DBR) stacks, which fundamentally alters their frequency response [9]. Finally, even in the evaluated back-to-back configuration utilizing a ~5 m Multimode Fiber (MMF) patch cord, coupling a multi-mode VCSEL into an MMF introduces non-negligible modal noise and modal dispersion. Given the short symbol duration at 53.125 Gbaud, these modal effects and the resulting inter-symbol interference become significant. Additionally, the VCSEL module operates without a thermoelectric cooler (TEC) at an ambient room temperature of ~20°C. This uncooled operation introduces dynamic self-heating and pronounced thermal transients. These specific hardware imperfections and thermal transients are entirely absent in standard single-mode DFB/SSMF setups, thereby necessitating the deployment of our advanced data-driven Bi-LSTM architecture to accurately capture the true system response.

## 3. Bi-LSTM Emulator Architecture

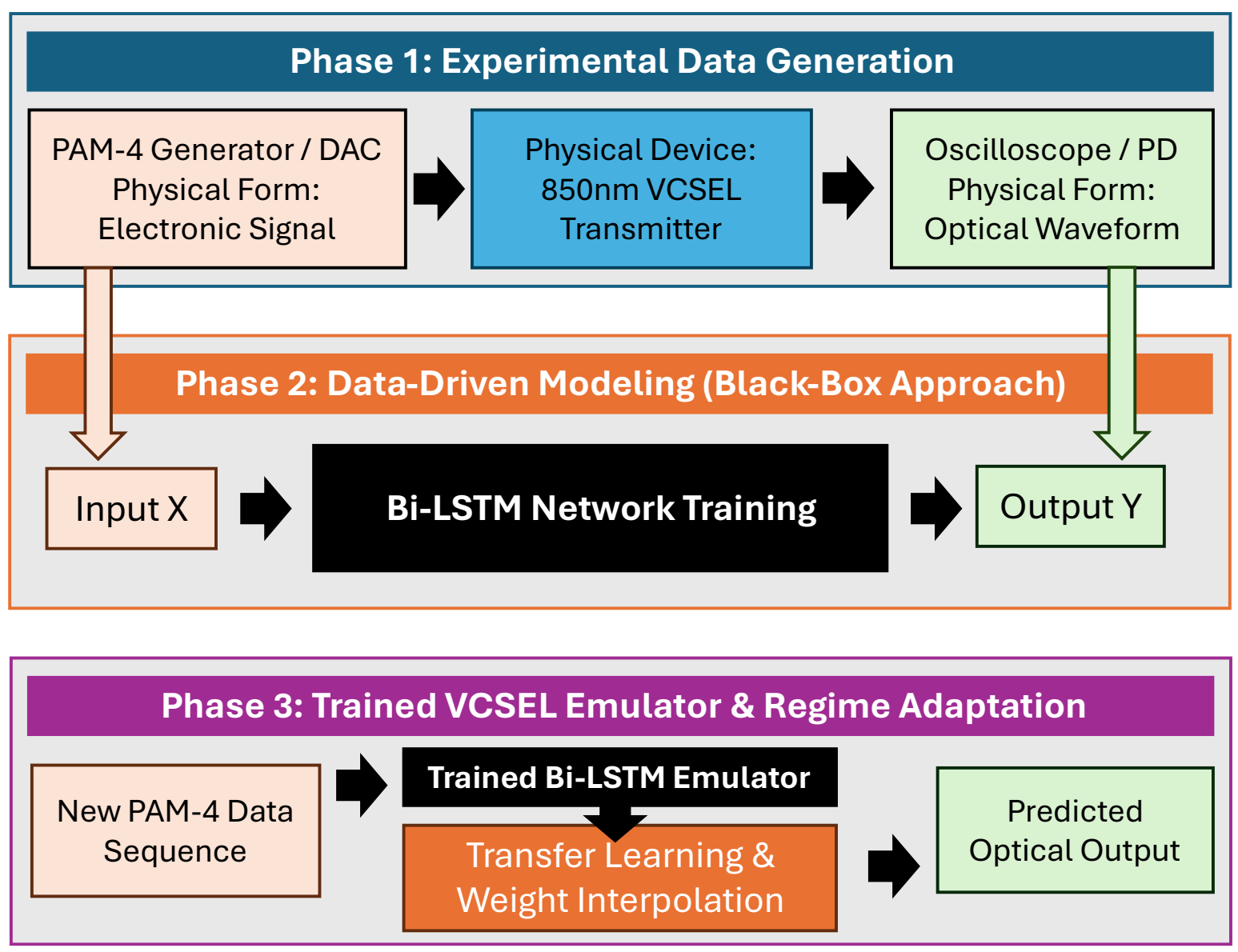

Fig. 2 Macroscopic data flow and end-to-end methodology of the proposed emulator. The flowchart illustrates the transition from experimental data generation to data-driven black-box training, and highlights the final emulation phase, which incorporates transfer learning and weight interpolation for rapid adaptation to new VCSEL operating regimes

To elucidate the macroscopic data flow and the end-to-end methodology of the proposed emulator, the system architecture is conceptually divided into three distinct phases, as illustrated in Fig. 2. In Phase 1 (Experimental Data Generation), the physical electronic PAM-4 signal (Input X) is generated by the DAC and drives the physical 850 nm VCSEL transmitter, producing the experimental optical waveform (Output Y) captured by the oscilloscope. In Phase 2 (Data-Driven Modeling), these corresponding electronic-in and optical-out data pairs are utilized to train the Bi-LSTM network via a black-box system identification approach. Finally, in Phase 3 (Emulation and Regime Adaptation), the fully trained Bi-LSTM acts as a standalone VCSEL emulator. When provided with a new, unseen electronic PAM-4 sequence, it analytically predicts the corresponding optical waveform. Crucially, as depicted in the augmented flowchart, this phase also incorporates our transfer learning and weight interpolation framework. Instead of training a new model from scratch for different operating conditions, the base network weights are either dynamically fine-tuned or mathematically interpolated to rapidly generate highly accurate emulators for entirely new VCSEL bias regimes.

Recurrent Neural Networks (RNNs) are well-known for their ability to process sequences of time-dependent data, making them highly suitable for tasks involving temporal correlations and memory effects. This characteristic makes RNNs an excellent choice for emulating the dynamic behavior of VCSEL-based optical transmission systems. Unlike simpler recurrent architectures, LSTM networks incorporate specialized components such as forget, input, and output gates.

These gates enable LSTMs to effectively manage both long-term and short-term dependencies.

To rigorously map the network parameters to the physical transmission system, the input sequence is formally defined as $X = [x_1, x_2, \ldots, x_L]$, where $x_t$ represents the normalized electrical amplitude of the PAM-4 signal at time step $t$, and $L$=80 is the sequence length (one sample per symbol). Internally, the LSTM computes the cell state $c_t$, which acts as the long-term memory capturing persistent physical phenomena such as thermal transients. This state is updated analytically as $c_t = f_t * c_{t-1} + i_t * tanh(W_{xc}x_t + W_{hc}h_{t-1} + b_c)$ where $f_t$ and $i_t$ are the analytically defined forget and input gates, and $*$ denotes element-wise multiplication. The short-term hidden state is then derived as $h_t = o_t * tanh(c_t)$ where $o_t$ representing the output gate.

Given the nonlinear nature and memory effects inherent in the optical channel, the Bi-LSTM was employed. Its bidirectional configuration allows the model to simultaneously consider past and future information. Physically, the forward hidden states $\overrightarrow{h_t}$ capture causal memory effects like carrier depletion, while the backward hidden states $h_t$provide the anti-causal context necessary for accurately regressing the finite-bandwidth smoothing effects of the VCSEL. The Bi-LSTM network with 28 hidden units was selected as optimal. Smaller networks compromised regression accuracy, while larger models offered no significant improvements. It was configured in a many-to-many approach [14], At the final stage of the network, the concatenated hidden states $h_t^{conc} = \left[\overrightarrow{h_t}, \overleftarrow{h_t}\right]$are passed to a fully connected layer (FCL) with a single neuron to produce the predicted optical output waveform $Y$, as depicted in the architecture flowchart in Fig. 3.

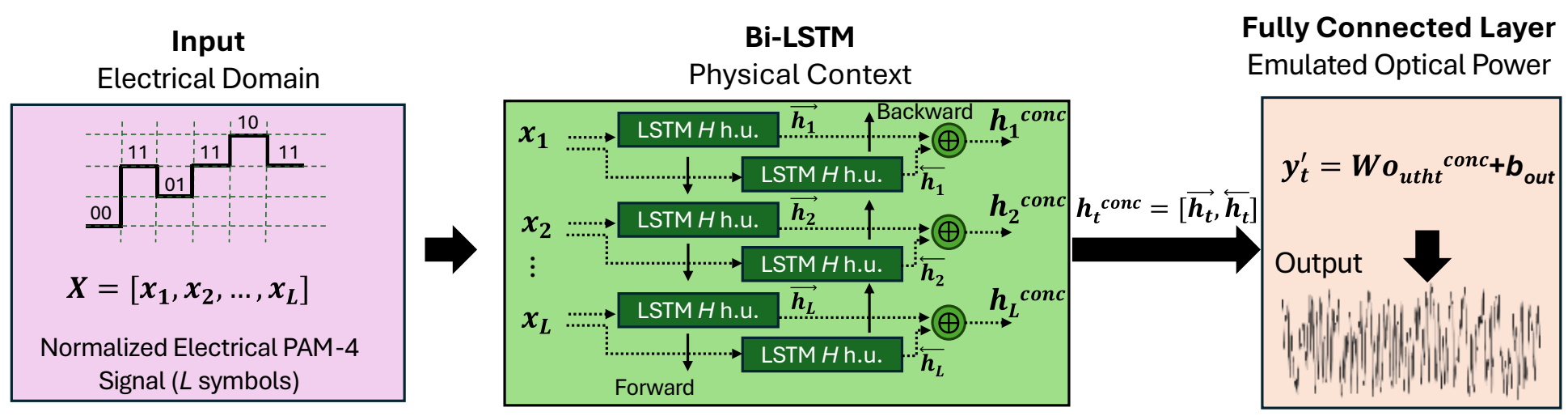

Fig.3 Architecture of the Bi-LSTM-based VCSEL emulator. The flowchart illustrates the processing of the input electrical sequence $X$ through the forward and backward LSTM layers, followed by the concatenation of their hidden states and the final FCL that predicts the optical output $Y$

The model was trained to predict the output of the VCSEL transmitter by taking the initial words (PAM-4 symbols) as input and learning from the experimentally recorded time series, accurately capturing the nonlinear behavior and memory effects inherent to the optical transmission system. For each operating regime of the VCSEL-based transmitter, two independent sequences of 1 million points each were recorded. However, after extensive testing, we found that using only 100,000 symbols for training provided the same emulation accuracy while offering a balanced trade-off between accuracy and training speed. Consequently, only 100,000 symbols from the first sequence were used for training, while the second sequence was entirely reserved for testing. The Bi-LSTM model was trained using 80% of the selected training data, with the remaining 20% allocated for validation. Training and testing were conducted in batches of 1000 words over a total of 1000 epochs using the TensorFlow framework with a GPU backend. The Adam optimization algorithm was employed to update the network weights, with a learning rate of $10^{-4}$ to ensure optimal convergence. The training was formulated as a regression task, utilizing the Mean Squared Error (MSE) as the loss function to minimize the difference between the predicted and actual output values. Analytically, the MSE is defined as:

$$\mathrm{MSE} = \frac{1}{K}\sum_{t=1}^{K}(y_t - \widehat{y_t})^2$$

where $y_t$ is the experimental ground truth (actual optical power) at time step $t$, $\widehat{y_t}$ is the network's predicted optical power, and $K$ is the total number of samples evaluated. Furthermore, to benchmark the overall emulation accuracy across different operating regimes, we utilized the Normalized Mean Squared Error (NMSE), which provides a scale-independent metric. The NMSE is calculated by normalizing the MSE with the variance of the experimental signal:

$$\mathrm{NMSE} = \frac{\sum_{t=1}^{K}(y_t - \widehat{y_t})^2}{\sum_{t=1}^{K}(y_t - \overline{y_t})^2}$$

where $\overline{y_t}$ is the mean value of the experimental target sequence.

To further enhance model generalization and mitigate overfitting, an early stopping mechanism with a patience of 50 epochs was implemented. This mechanism monitored the validation loss and terminated training if no improvement was observed for 50 consecutive epochs, ensuring an efficient and robust training process.

## 4. Performance Evaluation and Benchmarking

The dataset covers VCSEL operating regimes with electrical SNR above 20 dB, typical of modern short-reach IM/DD links. In the context of this study, an "operating regime" is strictly defined by the DC bias voltage applied to the VCSEL, while the peak-to-peak modulation amplitude, ambient operating temperature (~20°C, uncooled) and symbol rate (53.125 Gbaud) are kept constant. From a physical perspective, transitioning between these bias voltages fundamentally alters the VCSEL internal dynamics, specifically the relaxation oscillation frequency, damping factor, and steady-state carrier/photon densities, thereby altering the nonlinear distortion profile of the optical output. Under these high-SNR conditions, waveform distortions are dominated by VCSEL nonlinear dynamics, primarily large-signal amplitude compression and relaxation oscillations (AM-AM distortion), frequency chirp (AM-FM distortion), and bandwidth-related effects, rather than random noise. To rigorously demonstrate the inherent non-linearity of the system and the necessity of the proposed deep learning framework, we benchmarked the Bi-LSTM against a standard Linear Regression model trained on the exact same experimental dataset. As shown by the blue dotted line in Fig. 5, the Linear Regression model exhibits a significantly higher error (NMSE > 0.08) across all operating points. Crucially, the NMSE of the linear baseline distinctly increases as the bias voltage is raised. This degradation quantitatively confirms a physical transition to an increasingly nonlinear operating regime at higher voltages. In contrast, the Bi-LSTM emulator successfully

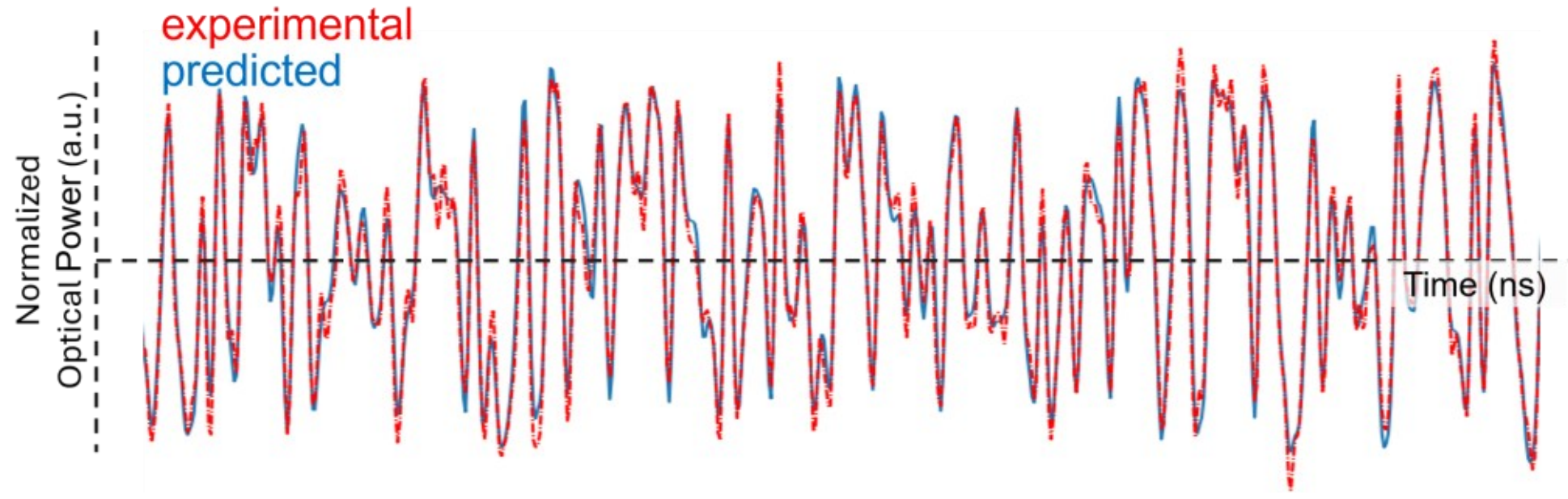

Fig. 4 Experimental (red) versus predicted VCSEL dynamics for Bi-LSTM. The specific evaluation was performed at a bias voltage of 1.4 V, a symbol rate of 53.125 Gbaud, and an ambient room temperature of 20°C, achieving an NMSE of 0.02

learns and reproduces the complex evolution of these nonlinear dynamics, maintaining a highly accurate and stable NMSE across the entire voltage sweep.

While NMSE serves as a rigorous waveform-level metric, its system-level implications are equally critical. The NMSE is inversely proportional to the effective Signal-to-Noise and Distortion Ratio (SNDR). A visual inspection of the time-domain sequence, as illustrated in Fig. 4 for the 1.4 V bias regime, confirms that the Bi-LSTM prediction tightly tracks the experimental waveform, accurately capturing both the rapid transient responses and the severe amplitude fluctuations. The achieved NMSE range of 0.02 to 0.04 translates to an effective electrical SNR of approximately 17 dB to 14 dB. According to the 400G Ethernet specifications (IEEE 802.3bs), this represents a highly critical operational boundary for 53 Gbaud PAM-4 systems. Specifically, an electrical SNR of ~16.5 dB is required to achieve a pre-FEC bit error rate (BER) of $2.4\text{x}10^{-4}$, which is the threshold for standard KP4 Reed-Solomon Forward Error Correction (FEC). By maintaining the NMSE within these bounds, our emulator operates precisely at the limits of practical system viability, ensuring that the predicted waveforms are reliable for physical-layer system evaluations.

### *4.1 Transfer Learning and Weight Interpolation*

It should be noted that while the base model trained in this study exhibits robust sequence generalization, accurately predicting arbitrary, unseen PAM-4 symbol sequences, it is not expected to generalize "zero-shot" to fundamentally different modulation formats (e.g., On-Off Keying (OOK) or PAM-8) or drastically different symbol rates. Data-driven models inherently learn the specific nonlinear dynamic responses explicitly excited by the frequency content and amplitude transitions of their training signal. However, this highlights the exact utility of our proposed transfer learning framework. The Bi-LSTM core, having already learned the fundamental physical nonlinearities of the VCSEL, acts as a generalized foundation that can be rapidly fine-tuned to new formats or conditions using only a minimal, specific dataset.

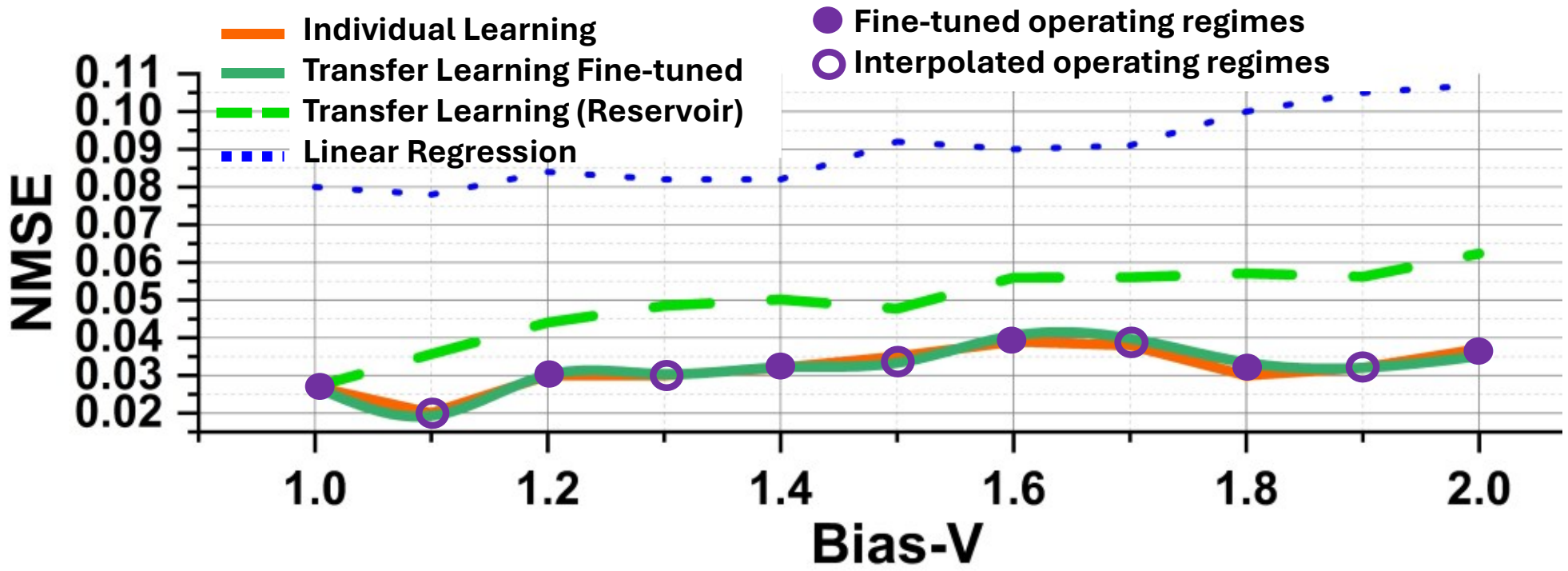

Fig. 5 Emulation accuracy (NMSE) across various bias voltages for Independent Learning and Transfer Learning approaches. The symbol rate (53.125 Gbaud) and ambient temperature (~20°C) were kept constant across all evaluated regimes. Solid lines are included to guide the eye. Filled markers denote fine-tuned operating regimes, while open markers denote interpolated operating points. The green dashed line represents the reservoir-like approach. The blue dotted line represents a baseline Linear Regression model, highlighting the inherent and increasing non-linearity of the system at higher bias voltages

In practice, the transfer learning process retains the entire Bi-LSTM architecture (28 hidden units and the fully connected layer), along with the fully converged weights of the base model initially trained at the 1.0 V regime. These pre-trained weights serve as the initialization state for the new operating condition, bypassing random initialization. The fine-tuning is performed over a dataset of 100,000 PAM-4 symbols obtained from the new regime. Consistent with the base training methodology, this dataset is split into 80% for fine-tuning and 20% for validation, and processed in batches of 1,000 words (where each word comprises 80 symbols). The network weights are dynamically updated via backpropagation using the Adam optimization algorithm with a constant learning rate of $10^{-4}$ and the Mean Squared Error (MSE) as the loss function. Because the model is already pre-conditioned to the underlying VCSEL dynamics, the early stopping mechanism, monitoring validation loss with a patience of 50 epochs, triggers significantly earlier than during a full training cycle. This rapid convergence terminates the fine-tuning process efficiently, confirming the substantial reduction in computational complexity.

To demonstrate this adaptability, transfer learning was employed to efficiently predict system behavior across multiple VCSEL operating regimes while reducing retraining time. A Bi-LSTM model initially trained at 1.0 V was incrementally adapted to new operating points (1.2 V, 1.4 V, 1.6 V, 1.8 V, and 2.0 V) using fine-tuning. Across ten regimes, the first-layer weights exhibited a nearly linear evolution between 1.0 V and 1.5 V, enabling accurate estimation at intermediate voltages (e.g., 1.1 V and 1.3 V) via linear interpolation of adjacent weight vectors. This unified approach eliminates the need for full retraining at each bias step and significantly reduces computational complexity.

In Fig. 5, fine-tuned operating points are shown with filled markers, while interpolated points are shown with open markers. The interpolated regimes achieve emulation accuracy equivalent to independently trained models, resulting in NMSE values closely matching those of direct training and demonstrating the robustness of combining transfer learning with weight interpolation.

### *4.2 Weight Sensitivity and Computational Efficiency*

To further analyze the significance of individual weight blocks within the Bi-LSTM network, Gaussian perturbations were applied directly to the weight blocks with a standard deviation $\sigma = 0.5\mu$, where $\mu$ denotes the mean absolute value of each block. This relative scaling follows established practices in weight-perturbation sensitivity analysis and provides a sufficiently strong yet non-destructive perturbation for revealing layer importance [15].

Table 1. Impact of gaussian noise on individual weight blocks in Bi-LSTM and their influence on prediction accuracy

| | NMSE |
|---|---|
| **Initial Weights** | **0,0415** |
| **noisy LSTM Forward 1x112** | **0,14 ± 0,08** |
| noisy LSTM Forward 28x112 | 0,10 ± 0,04 |
| noisy LSTM Backward 1x112 | 0,06 ± 0,02 |
| noisy LSTM Backward 28x112 | 0,043 ± 0,003 |
| **noisy Fully Connected 56x1** | **0,25 ± 0,12** |

The results, summarized in **Table 1**, indicate that the outermost weight blocks, specifically the initial input mapping block and the final Fully Connected layer, exert the most severe influence on regression accuracy. Conversely, the inner recurrent weights exhibit high robustness and minimal impact on the overall error.

This empirical finding directly justifies our "reservoir-like" transfer learning strategy. Unlike traditional reservoir computing where strictly only the output is trained, our approach freezes the entire inner recurrent core of the Bi-LSTM. This core, having already learned the fundamental memory effects and physical dynamics of the VCSEL at the base regime, acts as a robust nonlinear reservoir. During transfer learning to a new bias regime, we exclusively fine-tune the outermost boundary layers (both the initial input mapping and the final readout).

A key advantage of the proposed approach is its significant reduction in computational time during the deployment and inference phase, as summarized in Table 2. To benchmark this computational complexity, we utilized the standard coupled carrier-photon rate equations, solved via a 4th-order Runge-Kutta (RK4) numerical integration algorithm. Such traditional Ordinary Differential Equation (ODE) solvers rely on strictly sequential execution and require a stringent time step resolution. For a 53.125 Gbaud signal (symbol duration ~18.8 ps), accurately resolving the fast transient dynamics of the VCSEL, such as relaxation oscillations, necessitates a time step of $dt = 1$ ps. Consequently, the numerical solver is forced to calculate approximately 19 samples per symbol. Simulating 1 million PAM-4 symbols thus translates to processing nearly 19 million sequential samples, taking approximately 8.5 minutes per operating point on a standard Central Processing Unit (CPU). To sweep across the 6 evaluated bias regimes, this sequential process requires 51 minutes. In contrast, our deep learning approach operates directly at the target sampling rate and completes the initial training in 16 minutes, but by leveraging transfer learning, it adapts to new conditions in just 50 seconds per operating point. Further reductions are achieved through the reservoir-like training approach (fine-tuning strictly the boundary layers), which converges in just 35 seconds.

**Table 2. Comparison of Computational Efficiency for 1 Million PAM-4 Symbols (Note: Computation times for the Bi-LSTM and Transfer Learning approaches assume parallel execution on a GPU backend. The Rate Equations computation time assumes sequential execution on a standard CPU, as ODE solvers are fundamentally difficult to parallelize)**

| Modeling Approach | Computation Time | Scalability (Parameter Sweep) | Generalization Capability | NMSE |
|---|---|---|---|---|
| **Rate Equations (ODE)** | ~51 min (Sequential) | Low (Retune parameters) | Low (Requires refitting) | N/A (Theoretical) |
| **Bi-LSTM (Indep. Training)** | 16 min (Initial) | Medium | High | 0.02–0.04 |
| **Proposed Transfer Learning** | **< 1 min** | **High** | **High (Interpolation)** | **0.03–0.06** |

This emulator serves as a practical tool for optical engineers. By training on a sparse grid of experimental operating points (e.g., 5-6 bias voltages), the model can interpolate intermediate regimes almost instantaneously. This capability allows system designers to rapidly sweep through hundreds of potential bias and modulation configurations to identify optimal operating windows for BER minimization, a task that would be prohibitively slow using conventional rate-equation solvers.

## 5. Conclusion

This paper introduced a comprehensive data-driven framework utilizing Bi-LSTM networks for the highly efficient emulation of VCSEL-based PAM-4 transmission systems. Trained directly on experimental waveforms, the model successfully captured the complex device nonlinearities, achieving excellent prediction accuracy with NMSE values consistently bounded between 0.02 and 0.04 across diverse operating bias regimes. By leveraging transfer learning, the emulator adapts to new operating conditions in under a minute, drastically outperforming the computational bottlenecks of traditional rate-equation solvers. Furthermore, we demonstrated a reservoir-like training strategy that fine-tunes strictly the boundary layers, reducing adaptation time to just 35 seconds. Crucially, our study revealed that the network weights evolve linearly between fine-tuned states. This enables the instant generation of highly accurate models for intermediate operating regimes purely through mathematical weight interpolation, completely bypassing the need for additional experimental data or retraining. Ultimately, these advancements establish our deep learning framework as a highly practical and scalable rapid-prototyping tool for optical engineers designing next-generation intra-data center interconnects.

## Back matter

**Funding.** H.F.R.I. (Project Number: 28522, MALLORCA).

**Disclosures.** The authors declare no conflicts of interest.

**Data availability.** Data underlying the results presented in this paper are not publicly available at this time but may be obtained from the authors upon reasonable request.